\documentclass[showpacs,preprintnumbers,prb]{revtex4}
\usepackage{amssymb}
\usepackage{amsmath}
\usepackage{graphicx}
\usepackage{dcolumn}
\usepackage{bm}

\input{tcilatex}

\begin{document}

\title{Virgin magnetization of a magnetically shielded superconductor wire:
\\
theory and experiment}
\author{Yu.A.~Genenko}
\email{yugenen@tgm.tu-darmstadt.de}
\author{S.V.~Yampolskii}
\altaffiliation[On leave from ]{Donetsk Institute for Physics and Technology, National Academy of Sciences
of Ukraine, 83114 Donetsk, Ukraine.}
\affiliation{Institut f\"{u}r Materialwissenschaft, Technische Universit\"{a}t Darmstadt,
D-64287 Darmstadt, Germany}
\author{A.V.~Pan}
\affiliation{Institute for Superconducting and Electronic Materials, University of
Wollongong, Wollongong, NSW 2522, Australia}
\date{\today }

\begin{abstract}
On the basis of exact solutions to the London equation, the magnetic moment
of a type II superconductor filament surrounded by a soft-magnet environment
is calculated and the procedure of extracting the superconductor
contribution from magnetic measurements is suggested. Comparison of
theoretical results with experiments on MgB$_{2}/$Fe wires allows estimation
of the value of critical current for the first magnetic flux penetration.
\end{abstract}

\maketitle

Recently hybrid systems composed of superconducting and soft-magnetic
materials attracted much attention in view of possibilities to improve
superconductor performance by shielding out an external field as well as a
transport current self-field~\cite{Campbell,Genenko1,Jooss2}. Very intense
investigations were carried out on superconducting MgB$_{2}$ wires sheathed
in iron, which became ideal objects to explore the magnetic shielding effect
due to simplicity of their fabrication. As was observed in recent
experiments, such structures exhibit enhanced superconducting critical
currents over a wide range of the external magnetic field~\cite{Dou2,Dou3}.

The commonly used technique for estimation of the critical current value is
the measurement of superconductor magnetization versus applied magnetic
field. Usually such measurements are carried out as follows~\cite{Dou3,Dou4}%
. The total magnetization of MgB$_{2}/$Fe wire is measured in
superconducting state (below $T_{c}$) and in normal state (above $T_{c}$).
After that the magnetization of superconducting core is determined by
subtraction of the latter results from the former ones (because in the
normal state only magnetic sheath is magnetized). The magnetization of
superconductor allows to estimate the critical current value which is
proportional to the height of the hysteretic magnetic loop.\ It is assumed
in this procedure that the magnetization of iron sheath does not depend on
presence of superconductor and, hence, is identical above and below $T_{c}$.
However, it is intuitively clear that this assumption may be somewhat
incorrect. Indeed, due to the Meissner effect below $T_{c}$ the
superconductor expels the magnetic flux into the sheath. This expulsion does
not happen in the normal state where the magnetic field is homogeneous in
the cylindrical magnetically shielded cavity~\cite{Jackson}. Therefore, the
magnetic field distribution in the magnet sheath as well as its
magnetization can be different, depending on whether the core is in the
superconducting state or in the normal one. Recently, this scenario has been
supported by magneto-optical visualization of local flux distributions
within the iron sheath of MgB$_{2}$ superconducting wire \cite{Pan}.

In the present Letter we calculate exactly the distribution of magnetic
field inside and outside a superconducting filament sheathed by a magnet
layer, as well as the magnetization of such structure in the region of
reversible magnetic behavior, i.e. for the flux-free (Meissner) state of the
superconductor and well below the saturation field of the magnet. Comparing
theoretical results with experiment we verify the above described procedure
of the superconducting critical current estimation.

\begin{figure}[tbp]
\includegraphics{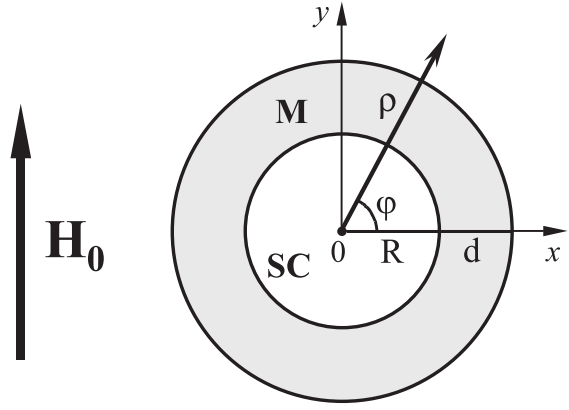}
\caption{Cross-sectional view of a superconductor filament covered
by a coaxial cylindrical magnetic sheath.} \label{fig1}
\end{figure}

Let us consider an infinite cylindrical superconductor filament of radius $R$
enveloped in a coaxial cylindrical magnetic sheath of thickness $d$ with
relative permeability $\mu $\ and exposed to the external magnetic field $%
\mathbf{H}_{0}$ perpendicular to the cylinder axis (Fig.~\ref{fig1}).

We start from the London equation for the magnetic induction $\mathbf{B}%
_{SC} $\ in the superconducting area~\cite{deGennes}%
\begin{equation}
\mathbf{B}_{SC}+\lambda ^{2}\func{curl}\func{curl}\mathbf{B}_{SC}=0,  \tag{1}
\label{1}
\end{equation}%
with the London penetration depth $\lambda $. The field outside the
superconductor denoted by $\mathbf{H}_{M}$\ in a magnetic sheath and by $%
\mathbf{H}_{out}$\ in a surrounding free space is described by the Maxwell
equations%
\begin{equation}
\func{curl}\mathbf{H}=0,\quad \func{div}\mathbf{H}=0,  \tag{2}  \label{2}
\end{equation}%
the latter of which is valid in the whole space. Implying an insulating,
nonmagnetic layer of thickness much less than $d$ and\ $R$ between the
superconductor and the magnet sheath~\cite{Dou5} the boundary conditions read%
\begin{align}
B_{n,SC}& =\mu _{0}\mu H_{n,M},\quad B_{t,SC}=\mu _{0}H_{t,M};  \tag{3a}
\label{3a} \\
\mu H_{n,M}& =H_{n,out},\quad \quad H_{t,M}=H_{t,out},  \tag{3b}  \label{3b}
\end{align}%
for the normal ($n$) and tangential ($t$) components on the
superconductor/magnet interface (Eq.~(\ref{3a})) and on the outer magnet
surface (Eq.~(\ref{3b})), respectively.
In addition, the field $\mathbf{H}%
_{out}$ has to approach asymptotically the external field $\mathbf{H}_{0}$.
In cylindrical coordinates ($\rho ,\varphi ,z$) coaxial with the filament
the solution of Eqs.~(\ref{1}-\ref{2}) is:%
\begin{eqnarray}
B_{\rho ,SC} &=&\mu _{0}H_{0}A_{SC}\left[ I_{0}\left( \rho /\lambda \right)
-I_{2}\left( \rho /\lambda \right) \right] \sin \varphi ,  \TCItag{4}
\label{HSC} \\
B_{\varphi ,SC} &=&\mu _{0}H_{0}A_{SC}\left[ I_{0}\left( \rho /\lambda
\right) +I_{2}\left( \rho /\lambda \right) \right] \cos \varphi ,  \notag
\end{eqnarray}%
in the superconductor;%
\begin{eqnarray}
H_{\rho ,M} &=&H_{0}\left( A_{M1}-A_{M2}R^{2}/\rho ^{2}\right) \sin \varphi ,
\TCItag{5}  \label{HML} \\
H_{\varphi ,M} &=&H_{0}\left( A_{M1}+A_{M2}R^{2}/\rho ^{2}\right) \cos
\varphi ,  \notag
\end{eqnarray}%
in the magnet sheath; and%
\begin{eqnarray}
H_{\rho ,out} &=&H_{0}\left[ 1+A_{out}\left( R+d\right) ^{2}/\rho ^{2}\right]
\sin \varphi ,  \TCItag{6}  \label{Hspace} \\
H_{\varphi ,out} &=&H_{0}\left[ 1-A_{out}\left( R+d\right) ^{2}/\rho ^{2}%
\right] \cos \varphi ,  \notag
\end{eqnarray}%
in the space around the filament. The coefficients $A_{SC},$ $A_{M1},$ $%
A_{M2}$\ and $A_{out}$ are given by expressions%
\begin{eqnarray}
A_{SC} &=&4\mu /\Delta ,  \TCItag{7}  \label{A} \\
A_{M1} &=&2\left[ \left( \mu +1\right) I_{0}\left( R/\lambda \right) +\left(
\mu -1\right) I_{2}\left( R/\lambda \right) \right] /\Delta ,  \notag \\
A_{M2} &=&2\left[ \left( \mu -1\right) I_{0}\left( R/\lambda \right) +\left(
\mu +1\right) I_{2}\left( R/\lambda \right) \right] /\Delta ,  \notag \\
A_{out} &=&\left\{ \left[ \left( \mu -1\right) ^{2}-\left( \mu +1\right)
^{2}R^{2}/\left( R+d\right) ^{2}\right] I_{2}\left( R/\lambda \right) \right.
\notag \\
&&\left. +\left( \mu ^{2}-1\right) \left[ 1-R^{2}/\left( R+d\right) ^{2}%
\right] I_{0}\left( R/\lambda \right) \right\} /\Delta ,  \notag
\end{eqnarray}%
where%
\begin{eqnarray}
\Delta &=&\left[ \left( \mu +1\right) ^{2}-\left( \mu -1\right)
^{2}R^{2}/\left( R+d\right) ^{2}\right] I_{0}\left( R/\lambda \right)
\TCItag{8}  \label{det} \\
&&+\left( \mu ^{2}-1\right) \left[ 1-R^{2}/\left( R+d\right) ^{2}\right]
I_{2}\left( R/\lambda \right) .  \notag
\end{eqnarray}%
The Meissner current density in the superconductor has only $z$-component
which equals%
\begin{equation}
j_{z}\left( \rho ,\varphi \right) =A_{SC}\frac{2H_{0}}{\lambda }I_{1}\left(
\rho /\lambda \right) \cos \varphi .  \tag{9}  \label{9}
\end{equation}%
A limiting case of the hollow magnetic cylinder may be obtained from
Eqs.~(4-8) by setting $\lambda \rightarrow \infty $\ which results in a
nonzero homogeneous field inside the hole as expected from Ref.~\cite%
{Jackson}.

Now we can easily calculate the mean magnetization of both superconducting
core and iron sheath which, due to the geometry of the problem, has only $y$%
-component. The magnetization of the superconductor is%
\begin{equation}
M_{SC}=\frac{1}{V_{SC}}\int\limits_{V_{SC}}dV\left[ \overrightarrow{\rho }%
\times \overrightarrow{j}\right] _{y}=-2A_{SC}H_{0}I_{2}\left( R/\lambda
\right) ,  \tag{10}  \label{10}
\end{equation}%
where the factor 2\ is due to account for the far ends of the sample~\cite%
{Brandt}. The magnetization of the iron sheath is%
\begin{equation}
M_{M}=\frac{\mu -1}{V_{M}}\int\limits_{V_{M}}dVH_{y,M}=\left( \mu -1\right)
H_{0}A_{M1}.  \tag{11}  \label{11}
\end{equation}%
In the practically interesting case $R\gg \lambda $\ they become%
\begin{equation}
M_{SC}\simeq -\frac{4H_{0}}{\mu +1-\left( \mu -1\right) /\left( 1+d/R\right)
^{2}},  \tag{12}  \label{12}
\end{equation}

\begin{equation}
M_{M}\simeq \frac{2\left( \mu -1\right) H_{0}}{\mu +1-\left( \mu -1\right)
/\left( 1+d/R\right) ^{2}}.  \tag{13}  \label{13}
\end{equation}%
Although the magnetization of the sheath~(\ref{13}) does not contain $%
\lambda $\thinspace\ it does not coincide with the magnetization of a hollow
magnetic cylinder which can be obtained from Eq.~(\ref{11}) by setting $%
\lambda \rightarrow \infty $:%
\begin{equation}
M_{HC}\simeq \frac{2\left( \mu ^{2}-1\right) H_{0}}{\left( \mu +1\right)
^{2}-\left( \mu -1\right) ^{2}/\left( 1+d/R\right) ^{2}}.  \tag{14}
\label{14}
\end{equation}%
For the ratio of these two quantities an inequality%
\begin{equation}
\frac{M_{M}}{M_{HC}}=\frac{\left( \mu +1\right) ^{2}-\left( \mu -1\right)
^{2}/\left( 1+d/R\right) ^{2}}{\left( \mu +1\right) ^{2}-\left( \mu
^{2}-1\right) /\left( 1+d/R\right) ^{2}}>1  \tag{15}  \label{15}
\end{equation}%
holds which means the magnetic flux density increase due to the flux
expelled from the superconductor.

Although in the critical state of superconductors only part of the magnetic
flux is expelled, the same inequality $M_{M}>M_{HC}$\ should be still valid.
Therefore, we conclude that in previous considerations~\cite{Dou2,Dou3,Dou4}
the magnetization of the sheath below $T_{c}$ could be underestimated and,
hence, the magnetization of a superconductor together with the critical
current value could be underestimated too.

Note that slopes of both magnetizations~(\ref{13}) and (\ref{14}), $%
M_{M\left( HC\right) }/H_{0}$,\ have \textit{finite} values at any fixed
parameter $d/R$ even in the limit of $\mu \rightarrow \infty $. It differs
from the case of the field parallel to the filament axis, when the shielding
effect is absent for an infinitely long cylinder sheath \cite{Jackson} and
the corresponding slope $M_{M\left( HC\right) }/H_{0}=\mu -1$ rises
unbounded with $\mu $. In fact, the shielding effect of a much smaller
amplitude was also observed in the longitudinal geometry due to the finite
length of samples \cite{Dou2,Dou5}.

Now we use the theoretical expressions to fit experimental results. Details
of the sample preparation and measurement technique were given in Ref.~\cite%
{Dou3}. The sample had the length of $L=4.1$~mm, the radius of the
superconducting core $R=0.5$~mm, the magnet sheath thickness $d=0.25$~mm,
and the permeability $\mu =46$ measured in the longitudinal field~\cite{Dou5}%
. The difference between magnetic moments measured in the normal and in the
superconducting states was quite small (about 1\%) at fields $B_{0}=\mu
_{0}H_{0}<0.1$~T (see Fig.~2).
\begin{figure}[tp]
\includegraphics{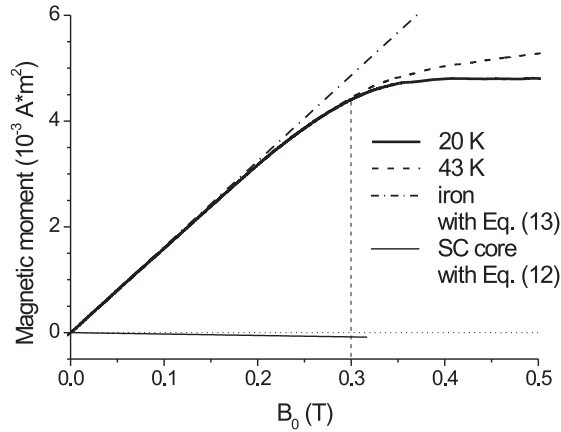}
\caption{The magnetic moment dependence of the MgB$_{2}$/Fe wire on the
magnetic field applied transversely in the superconducting state (thick
solid line) and in the normal state (dashed line). The moment of the iron
sheath in the superconducting state calculated using Eq.~(\protect\ref{13})
is presented by the dash-dotted line. The thin solid line exhibits the
superconductor core contribution.}
\label{fig2}
\end{figure}
The magnetic moment in the normal state was fitted using $M_{HC}$~(\ref{14})
while the total moment below $T_{c}$ was fitted using $M_{SC}$~(\ref{12})\
and $M_{M}$~(\ref{13})\ with the only fitting parameter of ratio $\eta $\
between the effective lengths of magnetic sheath and superconductor. The
best fit was achieved for $\eta =1.5$. The deviation of $\eta $ from $1$
indicates the significant role of the sample edge effects leading to the
discrepancy between the measured magnetic moment of the relatively short
sample and the moment calculated for an infinitely long cylinder with Eqs.~(%
\ref{12}-\ref{14}).

Using the above parameters we calculate the $M_{M}\left( B_{0}\right) $
dependence in the superconducting state from Eq.~(\ref{13}) which is shown
in Fig.~2 by the dash-dotted line. From Eqs.~(\ref{13}-\ref{14}) it follows
that the magnetization of the iron sheath at $T<T_{c}$ is about 3\% larger
than that at $T>T_{c}$. Let us note that this difference may reach 100\% at
smaller values of the parameters $d/R$ and $\mu $.

The response of the magnetic sheath is intrinsically nonlinear that is
clearly visible in the normal state magnetization curve at $B_{0}>0.15$~T.
The deviation of the total magnetic moment below $T_{c}$ at $B_{p}\simeq 0.3$%
~T from that in the normal state may be attributed to the nonlinearity due
to the first magnetic flux entry into the superconductor. From the $B_{p}$
value a critical current of the first vortex penetration may be estimated as
follows.

First, from the magnetization of the superconducting core the maximum value
of screening current $j_{s}=j_{z}\left( R,0\right) $ can be found. Combining
Eqs.~(\ref{9})-(\ref{10}) we obtain in the case $R\gg \lambda $%
\begin{equation}
j_{s}\simeq 2\left\vert M_{SC}\right\vert /\lambda ,  \tag{16}  \label{16}
\end{equation}%
with $M_{SC}$\ from Eq.~(\ref{12}). Taking $\mu =46$, $d/R=0.5$ and $\lambda
=1400$~\AA\ for $T=30$~K\ from Ref.~\cite{Finnemore} we obtain for $%
H_{0}=B_{p}/\mu _{0}$\ the value of $j_{s}\simeq 5\times 10^{7}$~A/cm$^{2}$.
A practically important quantity is the average density of the screening
current that may be defined as $j_{c}=2J_{c}/\pi R^{2}$ where $J_{c}$ is
determined by the integration of expression~(\ref{9}) over the half of the
superconductor cross-section. In the limit of $R\gg \lambda $\ we obtain%
\begin{equation}
j_{c}\simeq 8\left\vert M_{SC}\right\vert /\pi R,  \tag{17}  \label{17}
\end{equation}%
and for the parameters used we found $j_{c}\simeq 1.8\times 10^{4}$~A/cm$%
^{2} $ which is in a good agreement with the results of Refs.~\cite%
{Dou2,Dou3,Dou5}.

In conclusion, we have developed a procedure of extracting the
superconducting response from the low field magnetic measurements on the
iron sheathed superconductor filaments taking into account the difference
between the magnetization of the magnet sheath below and above $T_{c}$.%
\newline

Discussions of boundary conditions with A. Gurevich are gratefully
acknowledged. This study was supported by the Deutsche
Forschungsgemeinschaft (S.V.Y.) and by the Australian Research
Counsil (A.V.P.).

\bibliographystyle{plain}
\bibliography{apssamp}

\begin{thebibliography}{99}
\bibitem{Campbell} M.~Majoros, B.A.~Glowacki, and A.M.~Campbell, Physica C
\textbf{334}, 129 (2000); \textbf{338}, 251 (2000).

\bibitem{Genenko1} Yu.A.~Genenko, A.~Usoskin, and H.C.~Freyhardt, Phys. Rev.
Lett. \textbf{83}, 3045 (1999); Yu.A.~Genenko, A.~Snezhko, and
H.C.~Freyhardt, Phys. Rev. B \textbf{62}, 3453 (2000).

\bibitem{Jooss2} H.~Jarzina, Ch.~Jooss, and H.C.~Freyhardt, J.~Appl. Phys.
\textbf{91}, 3775 (2002).

\bibitem{Dou2} J.~Horvat, X.L.~Wang, S.~Soltanian, and S.X.~Dou, Appl. Phys.
Lett. \textbf{80}, 829 (2002).

\bibitem{Dou3} A.V.~Pan, S.H.~Zhou, H.K.~Liu, and S.X.~Dou, Supercond. Sci.
Technol. \textbf{16}, L33 (2003).

\bibitem{Dou4} M.D.~Sumption, E.W.~Collins, E.~Lee, X.L.~Wang, S.~Soltanian,
and S.X.~Dou, Physica C \textbf{378}, 894 (2002).

\bibitem{Jackson} J.D.~Jackson, \textit{Classical electrodynamics} (Wiley,
New York, 1975).

\bibitem{Pan} A.V. Pan, S.X. Dou, and T.H. Johansen, Proc. of NATO Advanced
Research Workshop on Magneto-Optical Imaging, {\O }ystese, Norway (2003) (in
press).

\bibitem{deGennes} P.G.~de Gennes, \textit{Superconductivity of Metals and
Alloys} (Addison-Wesley, New York, 1994).

\bibitem{Dou5} A.V.~Pan and S.X.~Dou (2003), unpublished.

\bibitem{Brandt} E.H.~Brandt and M.~Indenbom, Phys. Rev. B \textbf{48},
12893 (1993).

\bibitem{Finnemore} P.C.~Canfield, S.L.~Bud'ko, and D.K.~Finnemore, Physica
C \textbf{385}, 1 (2003).
\end{thebibliography}

\end{document}